\documentclass[10pt,a4paper,english]{article}
\usepackage[T1]{fontenc}
\usepackage[latin9]{inputenc}
\usepackage{color}
\usepackage{amsmath}
\usepackage{amssymb}
\usepackage{graphicx}
\makeatletter
\usepackage{amsfonts}
\usepackage{bbm}
\newcommand{\fif}{\varphi}
\newcommand{\eps}{\varepsilon}
\newcommand{\bra}[1]{\langle#1|}
\newcommand{\ket}[1]{|#1\rangle}
\newcommand{\cc}{\mathbb{C}}
\newcommand{\tr}{\mathrm{Tr}}

\newcommand{\bb}{\mathcal{B}}

\newcommand{\<}{\langle}
\renewcommand{\>}{\rangle}
\newcommand{\rot}{\mathcal{R}}
\newcommand{\pro}{\mathcal{P}}

\newcommand{\Mn}{M_n(\mathbb{C})}
\newcommand{\MM}{M_3(\mathbb{C})}

\newcommand{\ot}{{\,\otimes\,}}
\newcommand{{\Cd}}{{\mathbb{C}^d}}

\def\<{\langle}
\def\>{\rangle}

\newtheorem{theorem}{Theorem}

\newtheorem{proposition}{Proposition}
\newtheorem{example}{Example}

\newtheorem{conjecture}{Conjecture}

\begin{document}
\title{\textbf{Geometry of entanglement witnesses for two qutrits}}

\author{Dariusz Chru\'{s}ci\'{n}ski and Filip A. Wudarski\\
 Institute of Physics, Nicolaus Copernicus University,\\
 Grudzi\c{a}dzka 5/7, 87--100 Toru\'{n}, Poland}

\maketitle

\begin{abstract}
We characterize a convex subset of entanglement witnesses for two qutrits. Equivalently, we provide
a characterization of the set of positive maps in the matrix algebra of $3 \times 3$ complex matrices.
It turns out that boundary of this set displays elegant representation in terms of $SO(2)$ rotations. We conjecture that maps parameterized by rotations are optimal, i.e. they provide the strongest tool for detecting quantum entanglement.
As a byproduct we found a new class of decomposable entanglement witnesses parameterized by improper rotations from the orthogonal group $O(2)$.
\end{abstract}

\section{Introduction}

One of the most important problems of quantum information theory
\cite{QIT,HHHH,Guhne} is the characterization of mixed states of composed
quantum systems. In particular it is of primary importance to test
whether a given quantum state exhibits quantum correlation, i.e.
whether it is separable or entangled.

For low dimensional systems
there exists simple necessary and sufficient condition for
separability. The celebrated Peres-Horodecki criterium
\cite{Peres,PPT} states that a state of a bipartite system living in
$\mathbb{C}^2 \ot \mathbb{C}^2$ or $\mathbb{C}^2 \ot \mathbb{C}^3$
is separable iff its partial transpose is positive. Unfortunately,
for higher-dimensional systems there is no single universal
separability condition.

The most general approach to separability problem is based 
on the notion of an entanglement witness.  Recall, that a Hermitian  operator $W \in
\mathcal{B}(\mathcal{H}_A \ot \mathcal{H}_B)$ is an entanglement
witness \cite{Horodeccy-PM,Terhal1} iff: i) it is not positively
defined, i.e. $W \ngeq 0$, and ii) $\mbox{Tr}(W\sigma) \geq 0$ for
all separable states $\sigma$. A bipartite state $\rho$ living in
$\mathcal{H}_A \ot \mathcal{H}_B$ is entangled if and only if
there exists an entanglement witness $W$ detecting $\rho$, i.e. such that
$\mbox{Tr}(W\rho)<0$.

The separability problem may be equivalently formulated in terms of linear
positive maps: a linear map $\Phi : \mathcal{B}(\mathcal{H}_A) \longrightarrow
\mathcal{B}(\mathcal{H}_A)$ is positive if $\Phi(X) \geq 0 $ for all $X \geq 0$ from $\mathcal{B}(\mathcal{H}_A)$.
Now,  a bipartite state $\rho$ living in $\mathcal{H}_A \ot \mathcal{H}_B$ is separable if and only if
$({\rm id}_A \ot \Phi)\rho$ is positive for any positive map $\Phi$
from $\mathcal{H}_B$ into $\mathcal{H}_A$.  Positive maps play important role
both in physics and mathematics providing generalization of
$*$-homomorphism, Jordan homomorphism and conditional expectation.
Normalized positive maps define an affine mapping between sets of
states of $\mathbb{C}^*$-algebras. Unfortunately, in spite of the
considerable effort (see e.g.  \cite{Arveson}--\cite{Justyna3}), the
structure of positive maps (and hence also the set of entanglement
witnesses) is rather poorly understood.

In the present paper we analyze an important class of positive maps in $\MM$ introduced in \cite{Cho-Kye} ($\Mn$ denotes a set of $n \times n$ complex matrixes). This class provides natural generalization of positive maps in $\MM$ defined by Choi \cite{Choi}. Interestingly, the celebrated reduction map belongs to this class as well. We study the geometric structure of the corresponding  convex set. It turns out that part of its boundary defines an elegant class of positive maps parameterized by proper rotations from $SO(2)$. This class was already proposed in \cite{Kossak1} and generalized in \cite{kule}. Both Choi maps and reduction map corresponds to particular $SO(2)$ rotations. Equivalently, we provide the geometric analysis of the corresponding convex set of entanglement witnesses of two qutrits.

Interestingly, a convex set of positive maps displays elegant $\mathbb{Z}_2$--symmetry. We show that maps which are $\mathbb{Z}_2$--invariant are self-dual and decomposable. All remaining maps are indecomposable and hence may be used to detect bound entangled states of two qutrits. We conjecture that maps/entanglement witnesses  belonging to the boundary are optimal, i.e. they provide the strongest tool to detect quantum entanglement. This conjecture is supported by the following observations: i) both Choi maps and reduction map are optimal, ii) all maps from the boundary support another conjecture \cite{SPA1,SPA2} stating that so called structural physical approximation to an optimal entanglement witness defines a separable state. As a byproduct we constructed a new class of maps parameterized by improper rotations from $O(2)$. It is shown that all maps from this class are decomposable.

\section{A class of positive maps in $\MM$}

Let us consider a class of positive maps in $\MM$ defined as follows \cite{Cho-Kye}
\begin{equation}\label{MAPS}
    \Phi[a,b,c] = N_{abc} ( D[a,b,c] - {\rm id})\ ,
\end{equation}
where $D[a,b,c]$ is a completely positive linear map defined by
\begin{equation}\label{}
    D[a,b,c](X) = \left( \begin{array}{ccc} (a+1) x_{11} + bx_{22} + cx_{33} & 0 & 0 \\
    0 & cx_{11} + (a+1) x_{22} + bx_{33}   & 0 \\
    0 & 0 &  bx_{11} + cx_{22} + (a+1) x_{33}   \end{array} \right)\ ,
\end{equation}
with $x_{ij}$ being the matrix elements of $X \in \MM$, and `${\rm id}$' is an identity map, i.e. ${\rm id}(X)=X$ for any $X \in \MM$.
The normalization factor
\begin{equation}\label{}
    N_{abc} = \frac{1}{a+b+c} \ ,
\end{equation}
guarantees that $\Phi[a,b,c]$ is unital, i.e. $\Phi[a,b,c](\mathbb{I}_3) = \mathbb{I}_3$.
Note, that $N_{abc}D[a,b,c]$ is fully characterized by the following doubly stochastic circulant matrix
\begin{equation}\label{DS}
D= N_{abc}    \left(  \begin{array}{ccc} a & b & c \\ c & a &  b \\ b & c & a \end{array} \right)\ .
\end{equation}
This family contains well known examples of positive maps: note that $D[0,1,1](X) = \tr X\, \mathbb{I}_3$, and hence
\begin{equation}\label{R}
    \Phi[0,1,1](X) = \frac 12 ( \tr X\, \mathbb{I}_3 - X)\ ,
\end{equation}
which reproduces the reduction map. Moreover, $\Phi[1,1,0]$ and $\Phi[1,0,1]$   reproduce Choi map and  its dual, respectively \cite{Choi}. One proves the following result \cite{Cho-Kye}

\begin{theorem}  \label{TH-korea}
A map $\Phi[a,b,c]$ is positive but not completely positive if and only if
\begin{enumerate}
\item $0 \leq a < 2\ $,
\item $ a+b+c \geq 2\ $,
\item if $a \leq 1\ $, then $ \ bc \geq (1-a)^2$.
\end{enumerate}
Moreover, being positive it is indecomposable if and only if
\begin{equation}\label{ind}
  bc < \frac{(2-a)^2}{4}\ .
\end{equation}
\end{theorem}
Note, that for $a\geq 2$ the map $\Phi[a,b,c]$ is completely positive.  In this paper we analyze a class $\Phi[a,b,c]$
satisfying
\begin{equation}\label{abc=2}
    a+b+c=2 \ .
\end{equation}
Both reduction map (\ref{R}) and Choi maps belong to this class. It is clear that maps satisfying (\ref{abc=2})  belong to the boundary of the general class satisfying $a+b+c \geq 2$. Assuming (\ref{abc=2}) a family of maps (\ref{MAPS}) is essentially parameterized by two parameters
$$ \Phi[b,c]  := \Phi[2-b-c,b,c]\ . $$
Let us observe that condition {\em 3.} of Theorem \ref{TH-korea} defines a part of the boundary which corresponds to the part of the following ellipse
\begin{equation}\label{ellipse}
    \frac 94 \left(x - \frac 43 \right)^2 + \frac 34\, y^2 = 1\ ,
\end{equation}
where we  introduced new variables
$$  x= b+c \ , \ \ \ y=b-c \ , $$
Note, that condition for indecomposability (\ref{ind}) simplifies to $b \neq c$.
Hence, $\Phi[b,c]$ is decomposable iff $b=c$ which shows that decomposable maps lie on the line in $bc$--plane. This line intersects  the ellipse (\ref{ellipse}) in two points: $b=c=1$ which corresponds to the reduction map, and $b=c=1/3$.
A convex set of positive maps $\Phi[b,c]$ is represented
on the $bc$--plane on Figure~1.

\begin{figure}[htp] \label{fig}
 \centering
\includegraphics[scale=0.75]{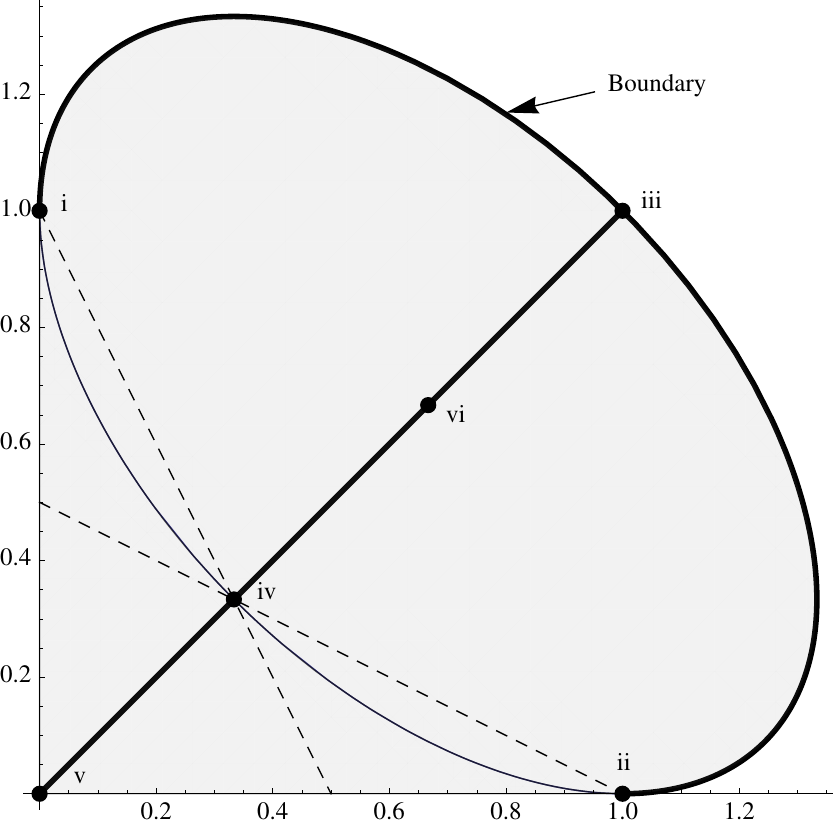}
\caption{A convex set of positive maps $\Phi[b,c]$. Red line $b=c$ corresponds to decomposable maps. Special points: (i) and (ii) Choi maps, (iii) reduction map, (v) is completely positive map, (iv) decomposable map with $b=c=1/3$.}
\end{figure}

Let us observe that this set is closed under simple permutation $(b,c) \rightarrow (c,b)$. Now, recall that for any map $\Phi : \MM \rightarrow \MM$ one defines its dual $\Phi^\# : \MM \rightarrow \MM$ by
$$  {\rm Tr}[ X \Phi(Y)] = {\rm Tr}[\Phi^\# (X) Y ]\ ,  $$
for all $X,Y \in \MM$. One easily finds
\begin{equation}\label{}
    \Phi^\# [b,c] = \Phi[c,b]\ ,
\end{equation}
that is, dual map to $\Phi[b,c]$ corresponds to permutation of $(b,c)$. This way we proved

\begin{proposition}
A map $\Phi[b,c]$ is decomposable if and only if it is self-dual.
\end{proposition}
The above class of positive maps gives rise to the class of entanglement witnesses
\begin{equation}\label{}
    W[a,b,c] = ({\rm id} \ot \Phi[a,b,c]) P^+\ ,
\end{equation}
where $P^+$ denotes a projector onto the maximally entangled state in $\mathbb{C}^3 \ot \mathbb{C}^3$. One finds the following matrix representation
\begin{equation}\label{}
  W[a,b,c]\, =\, \frac{N_{abc}}{3}\, \left( \begin{array}{ccc|ccc|ccc}
    a & \cdot & \cdot & \cdot & -1 & \cdot & \cdot & \cdot & -1 \\
    \cdot& b & \cdot & \cdot & \cdot& \cdot & \cdot & \cdot & \cdot  \\
    \cdot& \cdot & c & \cdot & \cdot & \cdot & \cdot & \cdot &\cdot   \\ \hline
    \cdot & \cdot & \cdot & c & \cdot & \cdot & \cdot & \cdot & \cdot \\
    -1 & \cdot & \cdot & \cdot & a & \cdot & \cdot & \cdot & -1  \\
    \cdot& \cdot & \cdot & \cdot & \cdot & b & \cdot & \cdot & \cdot  \\ \hline
    \cdot & \cdot & \cdot & \cdot& \cdot & \cdot & b & \cdot & \cdot \\
    \cdot& \cdot & \cdot & \cdot & \cdot& \cdot & \cdot & c & \cdot  \\
    -1 & \cdot& \cdot & \cdot & -1 & \cdot& \cdot & \cdot & a
     \end{array} \right)\ ,
\end{equation}
where to make the picture more transparent we replaced zeros by dots. Interestingly, all indecomposable witnesses $W[a,b,c]$ may be identified using the following family of PPT entangled (unnromalized) states:
\begin{equation}\label{eps}
    \rho_\epsilon = \sum_{i,j=1}^3 |ii\>\<jj| + \epsilon \sum_{i=1}^3 |i,i+1\>\<i,i+1| + \frac 1\epsilon \sum_{i=1}^3 |i,i+2\>\<i,i+2|\ ,
\end{equation}
where $\epsilon \in (0,\infty)$. It is well known that $\rho_\epsilon$ is PPT for all $\epsilon$  and entangled for $\epsilon \neq 1$. One easily finds
$$ {\rm Tr}\, (\rho_\epsilon W[a,b,c]) = N_{abc}\, \frac 1\epsilon\, ( b\epsilon^2 + [a-2]\epsilon + c) \ , $$
and hence $ {\rm Tr}\, (\rho_\epsilon W[a,b,c]) < 0$ might be satisfied only if the corresponding  discriminant
$$  (a-2)^2 - 4bc > 0\ , $$
which is equivalent to condition (\ref{ind}).

\section{A subclass parameterized by the rotation group}

Consider now positive maps $\Phi[b,c]$ belonging to the ellipse (\ref{ellipse}), i.e. satisfying $bc=(1-b-c)^2$. We show that these maps are uniquely characterized by the rotation group $SO(2)$. Let $f_\alpha$ ($\alpha=0,1,\ldots, n^2-1$) be an orthonormal basis in $\Mn$ such that  $f_0 = \frac{1}{\sqrt{n}}\, \mathbb{I}_n$, and $f_\alpha^*= f_\alpha$. One has
\begin{equation}\label{}
 {\rm Tr}\, ( f_k f_l) = \delta_{kl}\ , \ \ \ k,l=1,\ldots, n^2-1\ ,
\end{equation}
and ${\rm Tr} f_k = 0$ for $k=1,\ldots, n^2-1$. The following formula \cite{Kossak1} 
\begin{equation}\label{tw}
\Phi_R(X)=\frac{1}{n}\, \mathbb{I}_n \tr X + \frac{1}{n-1}\,\sum_{k,l=1}^{n^2-1} f_k R_{kl} \tr ( f_l X) \ ,
\end{equation}
where $R_{kl}$ is an  orthogonal matrix from $O(n^2-1)$,  defines a family of unital positive maps in $\Mn$ (for a slightly more general
construction see \cite{kule}).  It is not difficult to construct an orthonormal basis $f_\alpha$. One may take for example the generalized Gell-Mann matrices defined as follows: let $|1\>,\ldots,|n\>$ be an orthonormal basis in $\mathbb{C}^n$ and define
\begin{eqnarray*}
d_l &=& \frac{1}{\sqrt{l(l+1)}}\Big( \sum_{k=1}^l |k\>\<k|-l |l+1\>\<l+1|\Big)\ ,\ \ \ l=1,\ldots,n-1 \\
u_{kl} &=&\frac{1}{\sqrt{2}}(|k\>\<l|+|l\>\<k|)\ , \\
v_{kl} &=&\frac{-i}{\sqrt{2}}(|k\>\<l|-|l\>\<k|)\ ,
\end{eqnarray*}
for $k<l$. It is easy to see that $n^2$ Hermitian matrices $(f_0,d_l,u_{kl},v_{kl})$ define a proper orthonormal basis in $\Mn$.
Now, let us take $n=3$ and let
\begin{equation}\label{}
   R = \left(\begin{array}{c|c} T & 0 \\ \hline 0 & - \mathbb{I}_6 \end{array} \right)\ ,
\end{equation}
where $T \in O(2)$. An orthogonal group $O(2)$ has two connected components. Let us consider a proper rotation
\begin{equation}\label{T}
    T(\alpha) = \left(\begin{array}{cc} \cos\alpha & -\sin\alpha  \\ \sin\alpha & \cos\alpha \end{array} \right) \ ,
\end{equation}
for $\alpha \in [0,2\pi)$. It turns out that $\Phi[\alpha] := \Phi_R$ belongs to the class $\Phi[a,b,c]$. Indeed, one shows \cite{Kossak1} that $\alpha$-dependent coefficients $a,b,c$ are defined as follows
\begin{eqnarray}   \label{abc}
a(\alpha) &=&\frac{2}{3}\,(1+\cos\alpha)\ ,\nonumber\\
b(\alpha) &=&\frac{2}{3}\left(1-\frac{1}{2}\cos\alpha-\frac{\sqrt{3}}{2}\sin\alpha\right) \ ,\\
c(\alpha) &=&\frac{2}{3}\left(1-\frac{1}{2}\cos\alpha+\frac{\sqrt{3}}{2}\sin\alpha\right)\ ,\nonumber
\end{eqnarray}
and hence
$$ a(\alpha) + b(\alpha)+ c(\alpha)=2\ . $$
Now comes the crucial observation. It is easy to show that
\begin{equation}\label{ABC}
    b(\alpha)c(\alpha) = [1-a(\alpha)]^2\ ,
\end{equation}
for each $ \alpha \in [0,2\pi)$. Interestingly, one has
\begin{equation}\label{}
    a(\alpha)b(\alpha) = [1-c(\alpha)]^2\ , \ \ \ \
    a(\alpha)c(\alpha) = [1-b(\alpha)]^2\ ,
\end{equation}
that is, there is a perfect symmetry between parameters $(a,b,c)$. Hence, all maps $\Phi[\alpha]$ parameterized by $SO(2)$ belong to the characteristic ellipse (\ref{ellipse}) forming a part of the boundary of the simplex of $\Phi[b,c]$ (see Fig.~1). Note, that for $\alpha = \pm \pi/3$ one obtains two Choi maps ((i) and (ii) on Fig.~1), for $\alpha = \pi$ one obtains reduction map (point (iii) on Fig.~1) and for $\alpha=0$ one obtains decomposable map (point (iv) on Fig.~1).
Let us observe that $\Phi^\#[\alpha]= \Phi[-\alpha]$, and hence $\Phi[\alpha]$ is self-dual if and only if $\alpha=0$ or $\alpha=\pi$. The map $\alpha \rightarrow - \alpha$ realizes $\mathbb{Z}_2$ symmetry of our class of maps. Self-dual maps are $\mathbb{Z}_2$--invariant.

\section{Structural physical approximation}

It is well known that three points from the part of the boundary formed by the ellipse (\ref{ellipse}) define optimal positive maps (optimal entanglement witnesses): $(1,0)$ and $(0,1)$ corresponding to Choi maps, and $(1,1)$ corresponding to the reduction map.
In terms of $\Phi[\alpha]$ they correspond to $\alpha=\frac\pi 3,\frac{5}{3\pi}$ and $\alpha=\pi$, respectively.

Now,  for any  entanglement witness $W$ in $\mathcal{H}_A \ot \mathcal{H}_B$ such that ${\rm Tr}\, W=1$, one defines its structural physical approximation (SPA)
\begin{equation}\label{}
    \mathbf{W}(p) = (1-p)W + \frac{p}{d_Ad_B} \mathbb{I}_A \ot \mathbb{I}_B\ ,
\end{equation}
with $p \geq p^*$, where $p^*$ is the smallest value of $p$ such that $\mathbf{W}(p) \geq 0$.
Hence SPA of $W$ defines a legitimate quantum state $\mathbf{W}(p)$ in  $\mathcal{H}_A \ot \mathcal{H}_B$.
It was conjectured in \cite{SPA1} (se also recent paper \cite{SPA2}) that if $W$ is an optimal entanglement witness in $\mathcal{H}_A \ot \mathcal{H}_B$, then its SPA defines a separable state. This conjecture was supported by several examples of optimal entanglement witnesses (see e.g. \cite{Justyna1,Justyna2,Justyna3}). Now comes a natural question concerning optimality of other entanglement witnesses belonging to the boundary $\frac\pi 3 \leq \alpha \leq \frac{5}{3\pi}$. Let us recall a simple sufficient condition for optimality \cite{Lew}: if there exists a set  product vectors $|\psi \ot \phi\> \in \mathcal{H}_A \ot \mathcal{H}_B$ such that
$$ \< \psi \ot \phi|W|\psi \ot \phi\>=0\ , $$
and vectors $|\psi \ot \phi\>$ span the entire Hilbert space $\mathcal{H}_A \ot \mathcal{H}_B$, then $W$ is optimal.
Now, one can check that  $W[0,1,1]$ corresponding to $\alpha=\pi$ admits the full set (i.e. 9) of such vectors. For the rest points the problem is much more complicated \cite{Gniewko} (for $W[1,1,0]$ and $W[1,0,1]$ it was already shown in \cite{SPA1} that there are only 7 vectors). Nevertheless, as we show all these points supports the conjecture of \cite{SPA1}. We propose the following
\begin{conjecture} For $\frac\pi 3 \leq \alpha \leq \frac{5}{3\pi}$ positive maps $\Phi[\alpha]$ are optimal.
\end{conjecture}
Actually, it turns out that SPA for a large class of $W[a,b,c]$ defines a separable state. Let us consider
\begin{equation}\label{}
    \mathbf{W}(p) = (1-p)W[a,b,c] + \frac{p}{9}\, \mathbb{I}_3 \ot \mathbb{I}_3\ .
\end{equation}
Now, $\mathbf{W}(p) \geq 0$ for $p\geq p^*$, where the critical value $p^*$ is given by
\begin{equation}\label{}
    p^* = \frac{3(2-a)}{2 + 3(2-a)}\ .
\end{equation}
One easily finds 
\begin{equation}\label{}
    \mathbf{W}(p^*) = \frac{1}{3[2 + 3(2-a)]}\, \left\{ \sum_{i=1}^3 \Big( 2 |ii\>\<ii| + (2b+c) |i,i+1\>\<i,i+1| + (2c+b)|i,i+2\>\<i,i+2| \Big)  - \sum_{i\neq j} |ii\>\<jj|  \right\} \ ,
\end{equation}
where we have used $a+b+c=2$. Note, that $\mathbf{W}(p^*)$ may be decomposed as follows
\begin{equation}\label{}
     \mathbf{W}(p^*) = \frac{1}{3[2 + 3(2-a)]}\, \Big( \sigma_{12} + \sigma_{13} + \sigma_{23} + \sigma_d \Big) \ ,
\end{equation}
where
\begin{equation}\label{}
    \sigma_{ij} = |ij\>\<ij| + |ji\>\<ji| + |ii\>\<ii| + |jj\>\<jj| - |ii\>\<jj| - |jj\>\<ii| \ ,
\end{equation}
and the diagonal $\sigma_d$ reads as follows
\begin{equation}\label{}
    \sigma_d = \sum_{i=1}^3 \Big(  (2b+c-1) |i,i+1\>\<i,i+1| + (2c+b-1)|i,i+2\>\<i,i+2| \Big) \ .
\end{equation}
Now, $\sigma_{ij}$ are PPT and being supported on $\mathbb{C}^2 \ot \mathbb{C}^2$ they are separable. Clearly, $\sigma_d$ is separable whenever it defines a legitimate state, that is, $2b+c \geq 1$ and $2c + b \geq 1$. It defines a region in our simplex bounded by the part of the ellipse and two lines:
$$ c=1-2b\ , \ \ \ b = 1- 2c\ . $$
Interestingly, these lines intersect at $b=c= \frac 13$, i.e. point (iv) on Fig.~1.

\section{Decomposable maps parameterized by improper rotations}

Consider now a second component of $O(2)$ represented by the following family of matrices
\begin{equation}\label{T}
    \widetilde{T}(\alpha) = \left(\begin{array}{cc} \cos\alpha & \sin\alpha  \\ \sin\alpha & -\cos\alpha \end{array} \right) \ ,
\end{equation}
for $\alpha \in [0,2\pi)$.  Note, that ${\rm det}\, T(\alpha)=1$, whereas ${\rm det}\, \widetilde{T}(\alpha)=-1$.
One easily shows that in this case $\Phi[\alpha]$ leads to the following map
\begin{equation}\label{MAPS-tilde}
    \widetilde{\Phi}[a,b,c] = N_{abc} ( \widetilde{D}[a,b,c] - {\rm id})\ ,
\end{equation}
where $\widetilde{D}[a,b,c]$ is a completely positive linear map defined by
\begin{equation}\label{}
    \widetilde{D}[a,b,c](X) = \left( \begin{array}{ccc} (a+1) x_{11} + bx_{22} + cx_{33} & 0 & 0 \\
    0 & bx_{11} + (c+1) x_{22} + ax_{33}   & 0 \\
    0 & 0 &  cx_{11} + ax_{22} + (b+1) x_{33}   \end{array} \right)\ ,
\end{equation}
and $\alpha$-dependent coefficients $a,b,c$ are defined by
\begin{eqnarray}   \label{abc-new}
a(\alpha)&=&\frac{2}{3} \left(1+ \frac 12 \cos\alpha+\frac{\sqrt{3}}{2}\sin\alpha \right)\ ,\nonumber \\
b(\alpha)&=&\frac{2}{3}(1-\cos\alpha)\ ,\\
c(\alpha)&=& \frac{2}{3} \left( 1+  \frac 12 \cos\alpha-\frac{\sqrt{3}}{2}\sin\alpha \right)  \ . \nonumber
\end{eqnarray}
Note, that
$$ a(\alpha)+b(\alpha) + c(\alpha)=2\ , $$
and hence $\widetilde{D}[a,b,c]$ is fully characterized by the following  doubly stochastic  matrix
\begin{equation}\label{DS-tilde}
\widetilde{D} = \frac 12    \left(  \begin{array}{ccc} a & b & c \\ b & c &  a \\ c & a & b \end{array} \right)\ .
\end{equation}
Now, contrary to $D$ defined in (\ref{DS}) it is no longer circulant. Interestingly, new parameters (\ref{abc-new}) satisfy the same condition (\ref{ABC}) as $a,b,c$ defined in  (\ref{abc}), that is one has:
$$               b(\alpha)c(\alpha) = [1-a(\alpha)]^2\ , \ \ \ \         a(\alpha)b(\alpha) = [1-c(\alpha)]^2\ ,
\ \ \ \      a(\alpha)c(\alpha) = [1-b(\alpha)]^2\ . $$
It shows that $a,b,c$ defined in (\ref{abc-new}) belong to the same characteristic ellipse. It is therefore clear that points from the interior of this ellipse defines positive maps as well. This way we proved the following

\begin{theorem}
The linear map $\widetilde{\Phi}[a,b,c]$ defined by (\ref{MAPS-tilde}) with
\begin{itemize}

\item $ a,b,c \geq 0\ , $

\item $a+b+c=2\ , $

\item $  bc \geq (1-a)^2\ , $
\end{itemize}
is positive.
\end{theorem}
Equivalently, we constructed a new family of entanglement witnesses
\begin{equation}\label{ew1}
\widetilde{W}[a,b,c] = \frac 16 \left(\begin{array}{ccc|ccc|ccc}
a & \cdot & \cdot & \cdot & -1 & \cdot & \cdot & \cdot & -1 \\
\cdot & b & \cdot & \cdot & \cdot & \cdot & \cdot & \cdot & \cdot \\
\cdot & \cdot & c & \cdot & \cdot & \cdot & \cdot & \cdot & \cdot \\\hline
\cdot & \cdot & \cdot & b & \cdot & \cdot & \cdot & \cdot & \cdot \\
-1 & \cdot & \cdot & \cdot & c & \cdot & \cdot & \cdot & -1 \\
\cdot & \cdot & \cdot & \cdot & \cdot & a & \cdot & \cdot & \cdot \\\hline
 \cdot & \cdot & \cdot & \cdot & \cdot & \cdot & c & \cdot & \cdot \\
 \cdot & \cdot & \cdot & \cdot & \cdot & \cdot & \cdot &a & \cdot \\
 -1 & \cdot & \cdot & \cdot & -1 & \cdot & \cdot & \cdot & b
 \end{array}\right)\ .
\end{equation}
Let us observe that
\begin{equation}\label{}
    {\rm Tr}(\rho_\epsilon \widetilde{W}[a,b,c])= 0 \ ,
\end{equation}
where $\rho_\epsilon$ is defined in (\ref{eps}). Hence, this family of states does not detect indecomposability of $\widetilde{W}[a,b,c]$. Actually, one has the following

\begin{theorem}
All entanglement witnesses  $\widetilde{W}[a,b,c]$ are decomposable.
\end{theorem}
Proof: it is enough to prove this theorem from maps parameterized by points belonging to the ellipse $bc=(1-a)^2$, i.e. $a,b,c$ defined by (\ref{abc-new}). Note, that
\begin{equation}
\widetilde{W}[a,b,c] = \frac 16 \, (P + Q^\Gamma)\ ,
\end{equation}
where
\begin{equation*}
P=\left(\begin{array}{ccc|ccc|ccc}
a & \cdot & \cdot & \cdot & b-1 & \cdot & \cdot & \cdot & c-1 \\
\cdot & 0 & \cdot & \cdot & \cdot & \cdot & \cdot & \cdot & \cdot \\
\cdot & \cdot & 0 & \cdot & \cdot & \cdot & \cdot & \cdot & \cdot \\\hline
\cdot & \cdot & \cdot & 0 & \cdot & \cdot & \cdot & \cdot & \cdot \\
b-1 & \cdot & \cdot & \cdot & c & \cdot & \cdot & \cdot & a-1 \\
\cdot & \cdot & \cdot & \cdot & \cdot & 0 & \cdot & \cdot & \cdot \\\hline
 \cdot & \cdot & \cdot & \cdot & \cdot & \cdot & 0 & \cdot & \cdot \\
 \cdot & \cdot & \cdot & \cdot & \cdot & \cdot & \cdot &0 & \cdot \\
c -1 & \cdot & \cdot & \cdot & a-1 & \cdot & \cdot & \cdot & b
 \end{array}\right),\quad
 Q= \left(\begin{array}{ccc|ccc|ccc}
0 & \cdot & \cdot & \cdot & \cdot & \cdot & \cdot & \cdot & \cdot \\
\cdot & b & \cdot & -b & \cdot & \cdot & \cdot & \cdot & \cdot \\
\cdot & \cdot & c & \cdot & \cdot & \cdot & -c & \cdot & \cdot \\\hline
\cdot & -b & \cdot & b & \cdot & \cdot & \cdot & \cdot & \cdot \\
\cdot & \cdot & \cdot & \cdot & 0 & \cdot & \cdot & \cdot & \cdot \\
\cdot & \cdot & \cdot & \cdot & \cdot & a & \cdot & -a& \cdot \\\hline
 \cdot & \cdot & -c & \cdot & \cdot & \cdot & c & \cdot & \cdot \\
 \cdot & \cdot & \cdot & \cdot & \cdot & -a & \cdot &a & \cdot \\
 \cdot & \cdot & \cdot & \cdot & \cdot & \cdot & \cdot & \cdot & 0
 \end{array}\right) ,
\end{equation*}
and $Q^\Gamma$ denotes partial transposition of $Q$.
It is clear that $Q \geq 0$. Now, to prove that $P \geq 0$ let us observe  that the principal submatrix
\begin{equation}
\left(\begin{array}{ccc}a & b-1 & c-1 \\b-1 & c & a-1 \\c-1 & a-1 & b\end{array}\right)\ ,
\end{equation}
is positive semi-definite for $a,b,c$ defined by (\ref{abc-new}). Actually, the corresponding eigenvalues read $\{2,0,0\}$, i.e. they do not depend upon $\alpha$.
\hfill $\Box$

Let us observe that $W[a,b,c]= \widetilde{W}[a,b,c]$ if and only if $a=b=c=\frac 23$, that is, these two classes of entanglement witnesses have only one common element. Actually, this common point lies in the center of the ellipse, i.e. in the middle between point (iii) and (iv) on the Fig.~1.

Note, that entanglement witnesses $W[a,b,c]$ and $\widetilde{W}[a,b,c]$ differ by simple permutation along the diagonal. Let us define the following unitary matrix
\begin{equation}\label{}
    U = \left( \begin{array}{ccc} 1 & . & . \\ . & . & 1 \\ . & 1 & . \end{array} \right) \ ,
\end{equation}
which corresponds to permutation $(x,y,z) \rightarrow (x,z,y)$ and define
\begin{equation}\label{}
 {{W}}_U[a,b,c] :=   (U \ot \mathbb{I}_3) W[a,b,c]  (U \ot \mathbb{I}_3)^\dagger\ .
\end{equation}
Since $U \ot \mathbb{I}_3$ is a local unitary operator, ${W}_U[a,b,c]$ defines an entanglement witness.
 One easily finds
\begin{equation}\label{ew2}
{{W}}_U[a,b,c] =
 \frac 16 \left(\begin{array}{ccc|ccc|ccc}
a & \cdot & \cdot & \cdot & \cdot & -1 & \cdot & -1 & \cdot \\
\cdot & b & \cdot & \cdot & \cdot & \cdot & \cdot & \cdot & \cdot \\
\cdot & \cdot & c & \cdot & \cdot & \cdot & \cdot & \cdot & \cdot \\\hline
\cdot & \cdot & \cdot & b & \cdot & \cdot & \cdot & \cdot & \cdot \\
\cdot & \cdot & \cdot & \cdot & c & \cdot & \cdot & \cdot & \cdot \\
-1 & \cdot & \cdot & \cdot & \cdot & a & \cdot & -1 & \cdot \\\hline
 \cdot & \cdot & \cdot & \cdot & \cdot & \cdot & c & \cdot & \cdot \\
 -1 & \cdot & \cdot & \cdot & \cdot & -1 & \cdot &a & \cdot \\
 \cdot & \cdot & \cdot & \cdot & \cdot & \cdot & \cdot & \cdot & b
 \end{array}\right)\ ,
\end{equation}
which has the same diagonal as $\widetilde{W}[a,b,c]$ but the off-diagonal `$-1$' are distributed according to a different pattern.  We stress that ${{W}}_U[a,b,c]$ is an indecomposable entanglement witness for $b \neq c$, whereas $\widetilde{W}[a,b,c]$ is decomposable one.

\section{Conclusions}

We analyzed a geometric structure of the convex set of positive maps in $\MM$ (or equivalently a set of entanglement witness of
two qutrits). Interestingly, its boundary is characterized by proper rotations form $SO(2)$. It turns out that a positive map                                                                                                                            $\Phi[b,c]$ is decomposable if and only if it is self-dual. Hence maps which are not self-dual may be used as a tool for detecting PPT entangled states. As a byproduct we constructed a convex set of decomposable entanglement witnesses. The boundary of this set is now parameterized by improper rotations form $O(2)$. It is clear that a convex combination of $W[\alpha]$ and $\widetilde{W}[\beta]$ defines an entanglement witness as well. In particular, taking two probability distributions on a circle -- $p(\alpha)$ and $\widetilde{p}(\alpha)$ -- one defines a new class of entanglement witnesses
\begin{equation}\label{}
    W[p,\widetilde{p}] = \frac{1}{2\pi} \int_0^{2\pi} \Big( p(\alpha) W[\alpha] + \widetilde{p}(\alpha) \widetilde{W}[\alpha] \Big)d\alpha \ .
\end{equation}
Note however that mixing $W[\alpha]$ and $\widetilde{W}[\beta]$ we no longer control (in)decomposability of $W[p,\widetilde{p}]$ which strongly depends upon probability distributions $p$ and $\widetilde{p}$.

It would be interesting to generalize our analysis for $d>3$. The general case (even for $d=4$) is much more involved and the general structure of circulant entanglement witnesses is not known. Some results would be presented in a forthcoming paper.

\end{document}